\documentclass[12pt]{article}
\usepackage[cp1251]{inputenc}
\usepackage{amsmath}
\usepackage{amsfonts}
\usepackage{amssymb}
\def\pa{\partial}                       
\def\beq{\begin{eqnarray}}    
\def\eeq{\end{eqnarray}}      

\begin{document}
\date{}

\begin{center}
{\Large\textbf{On a gauge-invariant deformation\\ of a classical
gauge-invariant theory}}

\vspace{18mm}

{\large I.L. Buchbinder$^{(a,b)}\footnote{E-mail:
joseph@tspu.edu.ru}$\;,
P.M. Lavrov$^{(a,b,)} \footnote{E-mail:
lavrov@tspu.edu.ru}$,\;
}

\vspace{8mm}

\noindent  ${{}^{(a)}}
${\em
Center of Theoretical Physics, \\
Tomsk State Pedagogical University,\\
Kievskaya St.\ 60, 634061 Tomsk, Russia}

\noindent  ${{}^{(b)}}
${\em
National Research Tomsk State  University,\\
Lenin Av.\ 36, 634050 Tomsk, Russia}

\vspace{20mm}

\begin{abstract}
\noindent We consider a general gauge theory with independent
generators and study the problem of gauge-invariant deformation of
initial gauge-invariant classical action. The problem is formulated
in terms of BV-formalism and is reduced to describing the general
solution to the classical master equation. We show that such general
solution is determined by two arbitrary generating functions of the
initial fields. As a result, we construct in explicit form the
deformed action and the deformed gauge generators in terms of the
above functions. We argue that the deformed theory must in general
be non-local. The developed deformation procedure is applied to
Abelian vector field theory and we show that it allows to derive
non-Abelain Yang-Mills theory. This procedure is also applied to
free massless integer higher spin field theory and leads to local
cubic interaction vertex for such fields.

\end{abstract}

\end{center}

\vfill

\\

\newpage

\section{Introduction}
\noindent
Gauge theories are an integral part of the Standard Model
of Fundamental Interactions and the Standard Cosmological Model.
Therefore, going beyond the standard models can be related to the
construction and study of new gauge theories. In this paper, we
propose an approach to generate new gauge theories beginning with
some known and more or less simple gauge models. The approach is
based on Batalin-Vilkovisky (BV) formalism \cite{BV}, \cite{BV1},
\cite{BV2} that allows to explore a wide range of classical and
quantum aspects of the gauge theories by unified and universal way
(see the further development of the BV-formalism e.g. in \cite
{BLT-15}, \cite{BL-16}, \cite{BLT-21} and the references therein).

The BV-formalism was initially developed to provide a generic
universal method to quantize the general gauge theories. In this
paper, we study the purely classical problem of constructing a
deformation procedure of gauge theories with the preservation of
gauge invariance. It is worth noting here a certain analogy with the
use of the BRST-BFV method \cite{BFV1}, \cite{BFV2}, \cite{BFV3},
constructed initially for the covariant canonical quantization of
the gauge theories, in the classical higher-spin field theory (see
e.g. \cite{BPT-2001}, \cite{BK})\footnote{There is an extensive
literature devoted mainly to quantum aspects of BV-formalism. Since
we study the purely classical aspects, we are going to cite only the
papers which can in principle be related to our work.}.

The main object of the BV-formalism is the master equation which is
formulated in terms of the antibrackets \cite{BV}, \cite{BV1}. The
basic property of the antibracket is its invariance under the
anticanonical transformations. Namely this property plays an
important role in solving the different problems in classical and
quantum descriptions of the gauge systems. In this paper, we will use
the anticanonical transformations to find out the general solution
to the classical master equation that allows the construction of an arbitrary
gauge-invariant deformation of a given gauge theory.

The approach to solving the classical master equation was developed
in the papers \cite {BH}, \cite{H}, where it was proposed to look
for solutions to this equation in form of expansion in some coupling
parameters and reduce the classical master equation to an infinite
system of cohomologies\footnote{See also the recent papers \cite{D},
\cite{BaBu} and the references therein.}. Applications of this
approach to higher spin field theory were considered in the papers
\cite{BL}, \cite{BLS}, \cite{SS}. We develop a completely different
approach that does not require expansions and does not use the
cohomological analysis. The general solution of the classical master
equation is given in explicit form in terms of two independent
generating functions responsible for deformation of initial action
and initial gauge transformations.

The paper is organized as follows. In section 2 we briefly review
the basic notations of the BV-formalism such as the antibracket,
classical master equation, and anticanonical transformations.
Section 3 is devoted to the solution of the classical master
equation to construct the general deformation of the initial action
in terms of a single generating function depending on initial
fields. We prove that such deformation must in general be non-local,
although in some special cases the corresponding deformed action can
have the local sector. In section 4 we describe a general
deformation of gauge generators and derive the deformed gauge
algebra. The deformed generators are defined by two functions, one
of them is the same as for deformation of action and another
function relates to a special deformation of the generators. It is
interesting to point out that even if the initial theory is Abelian,
the deformed theory will obligatorily be non-Abelian. In section 5
we show that the application of the deformation procedure under
consideration to free Abelian vector field gauge theory leads to
deformed theory containing the standard non-Abelian Yang-Mills field
action among the other non-local terms in deformed action. Section 6
is devoted to the application of the above deformation theory to free
massless integer higher spin field theory that allows the
construction of the local cubic interaction vertex for such fields.
In section 7 we summarize the results.

In this paper, we systematically use the DeWitt's condensed notations
and employ the symbols $\varepsilon(A)$ for the Grassmann parity and
${\rm gh}(A)$ for the ghost number respectively. The right and left
functional derivatives are marked by special symbols $"\leftarrow"$
and $"\overrightarrow{}"$ respectively. Arguments of any functional
are enclosed in square brackets $[\;]$, and arguments of any
function are enclosed in parentheses, $(\;)$.

\section{Antibracket and master equation}

In this section, we briefly describe the basic notions of the
BV-formalism which will be essentially used in the paper to describe
a general gauge-invariant deformation of the classical gauge theory.

We consider a gauge theory of the fields $A=\{A^i\}$ with Grassmann
parities $\varepsilon(A^i)=\varepsilon_i$ and ghost numbers $\;{\rm
gh}(A^i)=0$. The theory is described by the initial action $S_0[A]$
and gauge generators $R^i{}_{\alpha}(A)$
($\varepsilon(R^i_{\alpha}(A))=\varepsilon_i+\varepsilon_{\alpha},\;
{\rm gh}(R^i_{\alpha}(A))=0$). The action is invariant under the
gauge transformations
\beq
\label{ggen0}
\delta A^i=R^i_{\alpha}(A)\xi^{\alpha},
\eeq
where the gauge parameters
$\xi^{\alpha}$ ($\varepsilon(\xi^{\alpha})=\varepsilon_{\alpha}$)
are the arbitrary functions of space-time coordinates. Condition of
gauge invariance is written in the standard form
\beq
\label{S0giv}
S_{0}[A]\overleftarrow{\pa}_{\!\!A^i}R^i_{\alpha}(A)=0.
\eeq
It is assumed that the fields $A=\{A^i\}$ are linear independent with
respect to the index $i$ however, in general, these generators may
be linear dependent with respect to index $\alpha$. Further, we
restrict ourselves by the irreducible gauge transformations with a
closed gauge algebra. In this case, the generators satisfy the
following relation
\beq
\label{ga} R^i_{\alpha ,
j}(A)R^j_{\beta}(A)-(-1)^{\varepsilon_{\alpha}\varepsilon_
{\beta}}R^i_{\beta ,j}(A)R^j_{\alpha}(A)=-R^i_{\gamma}(A)
F^{\gamma}_{\alpha\beta}(A),\quad R^i_{\alpha ,
j}(A)=R^i_{\alpha}(A)\overleftarrow{\pa}_{\!\!A^j},
\eeq
where
$F^{\gamma}_{\alpha\beta}(A)$
($\varepsilon(F^{\gamma}_{\alpha\beta}(A))=\varepsilon_{\alpha}+
\varepsilon_{\beta}+\varepsilon_{\gamma},\;{\rm
gh}(F^{\gamma}_{\alpha\beta}(A))=0$) are the structure coefficients
depending, in general, on the fields $A^i$ with the following
symmetry properties $F^{\gamma}_{\alpha\beta}(A)=
-(-1)^{{\varepsilon_{\alpha}\varepsilon_{\beta}}}F^{\gamma}_{\beta\alpha}
(A)$.

Following the BV-formalism, we introduce the minimal antisymplectic
space of fields $\phi^A$ and antifields $\phi^*_A$\footnote{We will
only be interested in possible deformations of initial classical
gauge systems consistent with basic properties of the BV-formalism.
In this case, it is sufficient to consider the minimal
antisymplectic space only.}
\beq
\phi^A=(A^i,C^{\alpha}),\quad
\phi^*_{A}=(A^*_i,C^*_{\alpha}),
\eeq
where $C^{\alpha}$
($\varepsilon(C^{\alpha})=\varepsilon_{\alpha}+1,\; {\rm
gh}(C^{\alpha})=1$) are the ghost fields and antifields obey the
following properties
\beq
\varepsilon(\phi^*_A)=\varepsilon(\phi^A)+1, \quad {\rm
gh}(\phi^*_A)=-1-{\rm gh}(\phi^A).
\eeq

The basic object of the BV-formalism is the extended action
$S=S[\phi,\phi^*]$ satisfying the classical master equation,
\beq
\label{master} (S,S)=0,
\eeq
and the boundary condition,
\beq
\label{boundary} S[\phi,\phi^*]\Big|_{\phi^*=0}=S_0[A].
\eeq
The master equation (\ref{master}) is written in terms of antibracket
which is defined for any functionals $F[\phi,\phi^*]$ and
$H[\phi,\phi^*]$ in the form
\beq
\label{antibracket} (G,H)=
G\left(\overleftarrow{\pa}_{\!\!\phi^A}\overrightarrow{\pa}_{\!\!\phi^*_{A}}-
\overleftarrow{\pa}_{\!\!\phi^*_{A}}\overrightarrow{\pa}_{\!\!\phi^A}\right)H.
\eeq

The gauge invariance of the initial action $S_0[A]$ leads to
invariance of the action $S[\phi,\phi^*]$,
\beq
\delta_B S=0
\eeq
under the global supersymmetry transformations (BRST transformations
\cite{brs1}, \cite{t})
\beq
\delta_B \phi^A=(\phi^A,S)\mu=
\overrightarrow{\pa}_{\!\!\phi^*_{A}}S\;\mu,\quad \delta_B
\phi^*_{A}=0,
\eeq
as a consequence the $S$ satisfies the classical
master equation. Here $\mu$ is a constant Grassmann parameter. In
the case of Yang-Mills theories, this invariance in the sector of
fields $A^i$ is nothing but the gauge invariance of $S_0[A]$ under
the standard gauge transformations with the special gauge parameters
$\xi^{\alpha}=C^{\alpha}\mu$.

Taking into account the gauge invariance of the initial action
(\ref{ggen0}) and the boundary condition (\ref{boundary}), one can
write the action $S=S[\phi,\phi^*]$ up to the terms linear in
antifields in the form \beq \label{S}
S=S_0[A]+A^*_iR^i_{\alpha}(A)C^{\alpha}-
\frac{1}{2}C^*_{\gamma}F^{\gamma}_{\alpha\beta}(A)C^{\beta}C^{\alpha}
(-1)^{\varepsilon_{\alpha}}+O(\phi^{*\;2}). \eeq We emphasize that
the antibracket is an essential element of the compact description
of the classical gauge theories within the BV-formalism. An
important property of the antibracket (\ref{antibracket}), which we
will use, is its invariance with respect to anticanonical
transformations of fields and anti-fields \cite{BV,BV1}. It leads to
the statement that any of the two solutions of classical master
equation (\ref{master}), satisfying the same boundary condition
(\ref{boundary}), are related one to another by some anticanonical
transformation \cite{BV,BV1}\footnote{It is worth pointing out that
the space of fields and antifields is analogous to phase space of
classical mechanics and the antibracket is analogous to Poisson
bracket. Then, the anticanonical transformations of the gauge theory
are analogous to canonical transformations preserving the Poisson
bracket.}. Moreover,  taking some solution of the classical master
equation satisfying the given boundary condition and using an
arbitrary anticanonical transformation in this solution, we will get
again a solution of the classical master equation. Namely this fact
will be used to construct a gauge-invariant deformation of the gauge
theories.

In this paper, we will describe the general deformation of initial
classical action and initial gauge symmetry on the base of the classical master equation.
We will see that these deformations are completely formulated in terms of
anticanonical transformations. It is known that there are two
possibilities to present the anticanonical transformations,
namely, in terms of generating functional or in terms of generators
\cite{BLT-15}. We suppose that for the problem under consideration, the description of
anticanonical transformations with the help of generating functional
seems to be more preferable.

\section{Deformed action}

In this section, we will describe the general structure of deformed action
depending on the same set of fields $A$ as the initial action $S_0[A].$

We begin with the action (\ref{S}) subjected to anticanonical
transformations
\beq \label{acS}
\widetilde{S}[\phi,\phi^*]=S[\Phi(\phi,\phi^*),
\Phi^*(\phi,\phi^*)],
\eeq
where $\Phi(\phi,\phi^*)$ and  $\Phi^*(\phi,\phi^*)$ are the solutions to the
equations
\beq
\phi^*_{A}=Y[\phi,\Phi^*]\overleftarrow{\pa}_{\!\!\phi^A},\quad
\Phi^A=\overrightarrow{\pa}_{\!\!\Phi^*_A}Y[\phi,\Phi^*],
\eeq
and
$Y=Y[\phi,\Phi^*]$ ($\varepsilon(Y)=1,\;{\rm gh}(Y)=-1$) is
the generating functional of the anticanonical transformation. The
action (\ref{acS}) satisfies the classical master equation,
\beq
\label{CMEacS} (\widetilde{S},\widetilde{S})=0
\eeq
and is invariant
under the BRST transformations,
\beq
\delta_B \widetilde{S}=0,\quad
\delta_B \phi^A=(\phi^A,\widetilde{S})\mu=
\overrightarrow{\pa}_{\!\!\phi^*_{A}}\widetilde{S}\;\mu,\quad
\delta_B \phi^*_{A}=0.
\eeq

To describe the possible deformations of the action, the generating
functional $Y$ should have the form
\beq
Y[\phi,\Phi^*]=\Phi^*_A\phi^A+X[\phi,\Phi^*],
\eeq
so that
\beq
\Phi^A=\phi^A+\overrightarrow{\pa}_{\!\!\Phi^*_A}X[\phi,\Phi^*],\quad
\phi^*_{A}=\Phi^*_A+X[\phi,\Phi^*]\overleftarrow{\pa}_{\!\!\phi^A}.
\eeq
Now let us consider the Taylor expansion of the functional $X$ in antifields,
\beq
X[\phi,\Phi^*]=\Phi^*_AH^A(\phi)+\frac{1}{2}\Phi^*_A\Phi^*_BH^{BA}(\phi)+
O(\phi^{*\;3}).
\eeq
Then we have
\beq \label{Ph}
&&\Phi^A=\phi^A+H^A(\phi)+\Phi^*_BH^{BA}(\phi)+O(\phi^{*\;2}),\\
&&
\label{Ph*}
\phi^*_{A}=\Phi^*_A+\Phi^*_BH^B(\phi)\overleftarrow{\pa}_{\!\!\phi^A}+
\frac{1}{2}\Phi^*_B\Phi^*_CH^{CB}(\phi)
\overleftarrow{\pa}_{\!\!\phi^A}+O(\phi^{*\;3}).
\eeq
First of all, we should express the $\Phi^*_A$ from (\ref{Ph*}) in the form
\beq
\label{sPh*}
\Phi^*_A=\Phi^*_A(\phi,\phi^*)
\eeq
and then to express $\Phi^A$ from (\ref{Ph})
as a function of variables $\phi$ and $\phi^*$,
\beq
\Phi^A(\phi,\phi^*)=\phi^A+H^A(\phi)+\Phi^*_B(\phi,\phi^*)H^{BA}(\phi)+
O(\phi^{*\;2}).
\eeq
The solution (\ref{sPh*}) can be found perturbatively.
The result reads
\beq
&&\Phi^A(\phi,\phi^*)\!=\!({\cal A}^{i}, {\cal C}^{\alpha})=
\phi^A+H^A(\phi)+\phi^*_C(M^{-1}(\phi))^C_{\;\;B}H^{BA}(\phi)+O(\phi^{*\;2}),\\
\label{sPh*1}
&&\Phi^{*}_A(\phi,\phi^*)\!=\!({\cal A}^*_{i}, {\cal C}^*_{\alpha})=
\phi^*_B(M^{-1}(\phi))^B_{\;\;A}\!+\!
\frac{1}{2}\phi^*_B\phi^*_CH^{CB}(\phi)
\overleftarrow{\pa}_{\!\!\phi^A}\!+\!O(\phi^{*\;3}),
\eeq
or, in a more detailed form,
\beq
\label{h}
&&{\cal A}^i=A^i+h^i(\phi)+\phi^*_A(M^{-1}(\phi))^A_{\;\;B}H^{Bi}(\phi)
+O(\phi^{*\;2}),
\eeq
\beq
\label{g}
&&{\cal C}^{\alpha}=C^{\alpha}+g^{\alpha}(\phi)+
\phi^*_A(M^{-1}(\phi))^{A}_{\;\;B}H^{B\alpha}(\phi)
+O(\phi^{*\;2})
\eeq
and
\beq
&&{\cal A}^*_i=A^*_j(M^{-1}(\phi))^{j}_{\;\;i}+
C^*_{\alpha}(M^{-1}(\phi))^{\alpha}_{\;\;i}+
+\frac{1}{2}\phi^*_B\phi^*_C H^{CB}(\phi)\overleftarrow{\pa}_{\!\!A^i}+
O(\phi^{*\;3}),\\
&&{\cal C}^*_{\alpha}=C^*_{\beta}(M^{-1}(\phi))^{\beta}_{\;\;\alpha}+
A^*_j(M^{-1}(\phi))^{j}_{\;\;\alpha}+
\frac{1}{2}\phi^*_B\phi^*_C H^{CB}(\phi)\overleftarrow{\pa}_{\!\!C^{\alpha}}+
O(\phi^{*\;3}),
\eeq
where $H^A(\phi)=(h^i(\phi),g^{\alpha}(\phi))$ and
 $(M^{-1}(\phi))^B_{\;\;A}$ is the inverse matrix for the matrix
\beq
\label{M}
M^B_{\;\;A}(\phi)=\delta^B_{\;\;A}+H^B(\phi)\overleftarrow{\pa}_{\!\!\phi^A}.
\eeq
The matrices $M^B_{\;\;A}(\phi)$ and $(M^{-1}(\phi))^B_{\;\;A}$
will be used to describe the deformation
of gauge symmetry.

From (\ref{sPh*1}) it follows the important relation
\beq
\label{result}
\Phi^*_A(\phi,\phi^*)\big|_{\phi^*=0}=0,
\eeq
which will be employed to study the possible deformations of action

If one uses the result (\ref{result}) in relation (\ref{acS}), one obtains
\beq
\label{dAc}
\widetilde{S}[\phi,\phi^*]\Big|_{\phi^*=0}=S[\Phi(\phi,\phi^*=0),0]=
S[\phi+H(\phi),0]=S_0[A+h(\phi)]
\eeq
with functions $h^i(\phi)$
defined by the expansion (\ref{h}).  This a final result for
deformation of the initial action $S[\phi,\phi^*]$ in the sector of
initial fields $A^i$. The result is extremely simple. The arbitrary
deformation of action is described  by a simple shift of the
field $A^i$ in the initial action $S_0[A]$ by arbitrary function
$h^i(\phi)$.

The result (\ref{dAc}) looks so simple that it can be considered as
trivial redefinition of the initial field $A^i$. Indeed, we can do
the inverse redefinition, exclude the functions $h^i(\phi)$ and
obtain a result that deformed theory is equivalent to initial theory.
However, such inverse redefinition leads to equivalent theory only
if the functions $h^i(\phi)$ are local. But when getting this
result, nowhere  it is assumed that these functions must be local. Hence
the nontrivial deformed action is obtained only for non-local
functions $h^i(\phi)$. Therefore we conclude that the deformed
action, in general, must be non-local functional. Nevertheless, this
does not mean that in some special cases a nontrivial deformed
action can not be local. In principle, there can be such a situation
that general non-local deformed action admits a closed local sector.
It means that the deformed action has the following structure
$S_0[A+h(A)] = S_1[A] + \textit{non-local terms},$ where the action
$S_1[A]$ is a local functional. If the deformed gauge
transformations contain a local piece that leaves the action
$S_1[A]$ invariant, we obtain the closed local sector of the deformed
theory. Let for example the initial Lagrangian has a form ${\cal L}
\sim A\Box A$ and let $h \sim \frac{1}{\Box}F(A)$ with some local
function $F(A).$ Then after some transformations, the deformed
Lagrangian takes the form $\tilde{{\cal L}} \sim A\Box A + 2F(A)+
\textit{non-local terms}.$ If the deformed gauge transformations
have a local piece leaving the Lagrangian  $\tilde{{\cal L}_1} \sim
A\Box A + 2F(A)$ invariant, we obtain the closed local sector of
the non-local theory. As we will see later, just such a situation is
realized to derive non-Abelian Yang-Mills theory from free
Abelian gauge theory. To conclude this section one notes that the
most general deformed action is obtained from the initial action
$S_0[A]$ by the transformation $A^i \rightarrow A^i+ h^i(A)$ with
non-local functions $h^i(A).$

\section{Deformed gauge symmetry}

In this section, we will describe the general structure of deformed gauge symmetry
of the  deformed action $S_0[A+h]$ as a direct consequence that the action
$\widetilde{S}=\widetilde{S}[\phi,\phi^*]$ satisfies the classical master equation.

\subsection{Consequences of the master equation}

To describe the deformation of gauge symmetry
of initial classical action we take into account that the action $\widetilde{S}$ satisfies the
master equation. Therefore it can be written in the form analogous to (\ref{S})
\beq
\widetilde{S}=S_0[{\cal A}] + {\cal A}^*_{i}R^i_{\alpha}({\cal A}){\cal C}^{\alpha}
-\frac{1}{2}{\cal C}^*_{\gamma}F^{\gamma}_{\alpha\beta}({\cal A})
{\cal C}^{\beta}{\cal C}^{\alpha}(-1)^{\varepsilon_{\alpha}}+O(\phi^{*\;2}),
\eeq
up to the second order in antifields $\phi^*_A$. Then from the classical master equation
for $\widetilde{S}$ one gets the relation
\beq
\label{ME*}
\frac{1}{2}(S_0[{\cal A}],S_0[{\cal A}])+(S_0[{\cal A}],{\cal A}^*_{i})\;
R^i_{\alpha}({\cal A}){\cal C}^{\alpha}
-\frac{1}{2}(S_0[{\cal A}],{\cal C}^*_{\gamma})\;F^{\gamma}_{\alpha\beta}({\cal A})
{\cal C}^{\beta}{\cal C}^{\alpha}(-1)^{\varepsilon_{\alpha}}+O(\phi^{*})=0.
\eeq
The terms in the l.h.s of (\ref{ME*}) can be represented in the forms
\beq
\label{terms}
&&\!\!\frac{1}{2}(S_0[{\cal A}],S_0[{\cal A}])=
S_0[{\cal A}]\overleftarrow{\pa}_{\!\!\phi^A}
(M^{-1}(\phi))^{A}_{\;\;B}H^{Bj}(\phi)
\overrightarrow{\pa}_{\!\!{\cal A}^j}S_0[{\cal A}]+O(\phi^{*}),\\
&&\!\!(S_0[{\cal A}],{\cal A}^*_{i})\;
R^i_{\alpha}({\cal A}){\cal C}^{\alpha}=
S_0[\widetilde{A}]\overleftarrow{\pa}_{\phi^A}(M^{-1}(\phi))^{A}_{\;\;i}
R^i_{\alpha}(\widetilde{A})\widetilde{C}^{\alpha}+O(\phi^{*}),\\
&&\!\!(S_0[{\cal A}],{\cal C}^*_{\gamma})\;F^{\gamma}_{\alpha\beta}({\cal A})
{\cal C}^{\beta}{\cal C}^{\alpha}(-1)^{\varepsilon_{\alpha}}\!=\!
S_0[\widetilde{A}]\overleftarrow{\pa}_{\!\!\phi^A}
(M^{-1}(\phi))^{A}_{\;\;\alpha}F^{\alpha}_{\sigma\beta}(\widetilde{A})
{\widetilde{C}}^{\beta}{\widetilde{C}}^{\sigma}(-1)^{\varepsilon_{\sigma}}
\!+\!O(\phi^{*}),
\eeq
where the notations
\beq
\label{h+g}
\widetilde{A}^i=A^i+h^i(\phi),\quad
\widetilde{C}^{\alpha}=C^{\alpha}+g^{\alpha}(\phi),
\eeq
are used. Here the functions $h^i$ and $g^{\alpha}$ are defined by the expansions (\ref{h}) and
(\ref{g}) respectively.

The first term in the r.h.s. (\ref{terms}) contains the functions $H^{Bj}$
which is defined by the
expansion (\ref{Ph}). To construct the deformed action
in the sector of initial fields,
we should switch off all antifields.
Therefore the functions $H^{Bj}$ do not enter the deformed action.
Hence, it is sufficient to put
\beq
H^{AB}(\phi)=0.
\eeq
Moreover, this restriction leads to the
correspondence
\beq
(S_0[A],S_0[A])=0 \quad \rightarrow \quad
(S_0[{\cal A}],S_0[{\cal A}])\big|_{\phi^*=0}=0.
\eeq
Now taking in the relation (\ref{ME*}) the limit $\phi^*\rightarrow 0,$
one obtains
\beq
S_0[\widetilde{A}]\overleftarrow{\pa}_{\phi^A}(M^{-1}(\phi))^{A}_{\;\;i}
R^i_{\alpha}(\widetilde{A})\widetilde{C}^{\alpha}
-\frac{1}{2}S_0[\widetilde{A}]\overleftarrow{\pa}_{\phi^A}
(M^{-1}(\phi))^{A}_{\;\;\alpha}
F^{\alpha}_{\sigma\beta}(\widetilde{A})
\widetilde{C}^{\beta}\widetilde{C}^{\sigma}(-1)^{\varepsilon_{\sigma}}=0.
\eeq

Since the initial fields $A^i$ obey the property ${\rm gh}(A^i)=0$,
the generating functions $h^i$ obey the analogous property ${\rm gh}(h^i)=0$.
Therefore,
the functions $h^i$ do not depend on the ghost fields $C^{\alpha}$
(${\rm gh}(C^{\alpha})=1$),
\beq
h^i(\phi)=h^i(A).
\eeq
It leads to the following relation
\beq
\label{SAg}
S_0[\widetilde{A}]\overleftarrow{\pa}_{A^j}(M^{-1}(\phi))^{j}_{\;\;i}
R^i_{\alpha}(\widetilde{A})\widetilde{C}^{\alpha}
-\frac{1}{2}S_0[\widetilde{A}]\overleftarrow{\pa}_{A^j}
(M^{-1}(\phi))^{j}_{\;\;\alpha}
F^{\alpha}_{\sigma\beta}(\widetilde{A})
\widetilde{C}^{\beta}\widetilde{C}^{\sigma}(-1)^{\varepsilon_{\sigma}}=0.
\eeq
The matrix $M^B_{\;\;A}(\phi)$ (\ref{M}) has the triangular  form
\beq
&&M^j_{\;\;i}(\phi)=\delta^j_{\;\;i}+h^j(A)\overleftarrow{\pa}_{\!\!A^i},\quad
M^{j}_{\;\;\beta}(\phi)=0,\\
&&M^{\alpha}_{\;\;i}(\phi)=g^{\alpha}(\phi)\overleftarrow{\pa}_{\!\!A^i}, \quad
\qquad\; M^{\alpha}_{\;\;\beta}(\phi)=\delta^{\alpha}_{\;\;\beta}+
g^{\alpha}(\phi)\overleftarrow{\pa}_{C^{\!\!\beta}}.
\eeq
It means that inverse matrix $(M^{-1}(\phi))^{B}_{\;\;A}$ has triangular
structure as well,
\beq
&& (M^{-1}(\phi))^{j}_{\;\;i},\quad (M^{-1}(\phi))^{j}_{\;\;\beta}=0,\\
&&(M^{-1}(\phi))^{\alpha}_{\;\;i},\quad (M^{-1}(\phi))^{\alpha}_{\;\;\beta},
\eeq
where $(M^{-1}(\phi))^{j}_{\;\;i}$ is inverse to $M^j_{\;\;i}(\phi)$,
\beq
(M^{-1}(\phi))^{j}_{\;\;l}M^l_{\;\;i}(\phi)=\delta^j_{\;\;i},
\eeq
and does not depend on fields $C^{\alpha}$,
\beq
(M^{-1}(\phi))^{j}_{\;\;i}=(M^{-1}(A))^{j}_{\;\;i}.
\eeq
In its turn $(M^{-1}(\phi))^{\alpha}_{\;\;\beta}$ is inverse to
$ M^{\alpha}_{\;\;\beta}(\phi)$,
\beq
(M^{-1}(\phi))^{\alpha}_{\;\;\gamma}M^{\gamma}_{\;\;\beta}(\phi)=
\delta^{\alpha}_{\;\;\beta}.
\eeq
As a consequence, the relation (\ref{SAg}) rewrites as
\beq
\label{SA0g}
S_0[\widetilde{A}]\overleftarrow{\pa}_{A^j}(M^{-1}(A))^{j}_{\;\;i}
R^i_{\alpha}(\widetilde{A})\widetilde{C}^{\alpha}=0.
\eeq
This relation is the base to derive the general deformation of gauge generators.

\subsection{Deformation of gauge generators}

We proceed with the discussion of transformed gauge symmetry. Since
${\rm gh}(g^{\alpha}(\phi))=1$ the
generating functions $g^{\alpha}(\phi)$ is linear in the ghost fields $C^{\alpha}$,
\beq
g^{\alpha}(\phi)=g^{\alpha}_{\;\;\beta}(A)C^{\beta}
\eeq
Therefore the relation (\ref{h+g}) leads to
\beq
\label{tildeC}
\widetilde{C}^{\alpha}=M^{\alpha}_{\;\;\beta}(A)C^{\beta}.
\eeq
Taking into account this relation, we rewrite the relation (\ref{SA0g}) in the form
\beq
\label{identity}
S_0[\widetilde{A}]\overleftarrow{\pa}_{A^j}(M^{-1}(A))^{j}_{\;\;i}
R^i_{\alpha}(\widetilde{A})M^{\alpha}_{\;\;\beta}(A)=0.
\eeq
Denoting
\beq
\label{generators}
{\bf R}^i_{\alpha}(A)=(M^{-1}(A))^{j}_{\;\;i}
R^i_{\alpha}(\widetilde{A})M^{\alpha}_{\;\;\beta}(A)
\eeq
and using the definition $S_0[\widetilde{A}]=S[A]$ one gets
\beq
\label{identity1}
S[A]\overleftarrow{\pa}_{\!\!A^j}{\bf R}^j_{\alpha}(A)=0.
\eeq
This relation allows us to interpret the ${\bf R}^i_{\alpha}(A)$ (\ref{generators})
as the deformed
gauge generators. Then the relation (\ref{identity1}) means
the condition of gauge invariance of the deformed action.

Let us now turn to the derivation of the gauge algebra for deformed
generators. To do that we should calculate the quantity
\beq
\label{lhs}
\big({\bf R}^i_{\alpha}(A)\overleftarrow{\pa}_{\!\!A^j}\big)
{\bf R}^j_{\beta}(A)-
(-1)^{\varepsilon_{\alpha}\varepsilon_
{\beta}}\big({\bf R}^i_{\beta}(A)\overleftarrow{\pa}_{\!\!A^j}\big)
{\bf R}^j_{\alpha}(A)
\eeq
The calculation is divided into few steps.

{\textbf{1}.} Let us write
\beq
\label{generators1}
{\bf R}^i_{\alpha}(A) = \widetilde{R}^i_{\beta}(A)M^{\beta}_{\;\;\alpha}(A)
\eeq
with
\beq
\label{generators2}
\widetilde{R}^i_{\beta}(A)= (M^{-1}(A))^{i}_{\!\;\;j}
R^j_{\alpha}(\widetilde{A}).
\eeq
Then the relation (\ref{S0giv}) can be represented in the form
\beq
\label{S0giv1}
S_{0}[\widetilde{A}]\overleftarrow{\pa}_{\!\!\widetilde{A}^i}
R^i_{\alpha}(\widetilde{A})=0,
\eeq
or, equivalently, as
\beq
\label{S0giv2}
S[A]\overleftarrow{\pa}_{\!\!A^i}
\widetilde{R}^i_{\alpha}(A)=0.
\eeq

{\textbf{2}.} Using the relation (\ref{S0giv1}) we rewrite the initial gauge identity
(\ref{ga}) in the form
\beq
\label{ga1}
\big(R^i_{\alpha}(\widetilde{A})\overleftarrow{\pa}_{\!\!\widetilde{A}^j}\big)
R^j_{\beta}(\widetilde{A})-
(-1)^{\varepsilon_{\alpha}\varepsilon_
{\beta}}\big(R^i_{\beta}(\widetilde{A})\overleftarrow{\pa}_{\!\!\widetilde{A}^j}\big)
R^j_{\alpha}(\widetilde{A})=
-R^i_{\gamma}(\widetilde{A})
F^{\gamma}_{\alpha\beta}(\widetilde{A}).
\eeq
Then one takes into account the relation
\beq
R^i_{\alpha}(\widetilde{A})=M^i_{\;j}(A)\widetilde{R}^i_{\alpha}(A)
\eeq
and properties of the matrix $M^i_{\;j}(A)$ (\ref{M})
\beq
M^i_{\;j}(A)\overleftarrow{\pa}_{\!\!A^k}=
h^i(A)\overleftarrow{\pa}_{\!\!A^j}\overleftarrow{\pa}_{\!\!A^k},\\
\nonumber
M^i_{\;j}(A)\overleftarrow{\pa}_{\!\!A^k}=
(-1)^{\varepsilon_j\varepsilon_k}M^i_{\;k}(A)\overleftarrow{\pa}_{\!\!A^j}.
\eeq
It allows us to rewrite the relation (\ref{ga1}) in the form
\beq
\label{ga2}
\big(\widetilde{R}^i_{\alpha}(A)\overleftarrow{\pa}_{\!\!A^j}\big)
\widetilde{R}^j_{\beta}(A)-
(-1)^{\varepsilon_{\alpha}\varepsilon_
{\beta}}\big(\widetilde{R}^i_{\beta}(A)\overleftarrow{\pa}_{\!\!A^j}\big)
\widetilde{R}^j_{\alpha}(A)=
-\widetilde{R}^i_{\gamma}(A)
\widetilde{F}^{\gamma}_{\alpha\beta}(A),
\eeq
where
$\widetilde{F}^{\gamma}_{\alpha\beta}(A)=F^{\gamma}_{\alpha\beta}(\widetilde{A})
$ and $F^{\gamma}_{\alpha\beta}(A)$ are the structure coefficients of initial gauge algebra.

{\textbf{3}.} Now one introduces the deformed generators (\ref{generators}) and use the
relations (\ref{ga2}) and explicit form of the matrix $M^{\alpha}_{\;\;\beta}(A)$
\beq
M^{\alpha}_{\;\beta}(A)=\delta^{\alpha}_{\;\beta}+g^{\alpha}_{\;\beta}(A).
\eeq
It allows us to transform the relation (\ref{ga2}) into the following relation
\beq
\label{gabold}
\big({\bf R}^i_{\alpha}(A)\overleftarrow{\pa}_{\!\!A^j}\big)
{\bf R}^j_{\beta}(A)-
(-1)^{\varepsilon_{\alpha}\varepsilon_
{\beta}}\big({\bf R}^i_{\beta}(A)\overleftarrow{\pa}_{\!\!A^j}\big)
{\bf R}^j_{\alpha}(A)=
-{\bf R}^i_{\gamma}(A)
{\bf F}^{\gamma}_{\alpha\beta}(A).
\eeq
Here
\beq
\label{newstructure}
{\bf F}^{\gamma}_{\alpha\beta}(A)&=&-(M^{-1})^{\gamma}_{\;\lambda}(A)
\widetilde{F}^{\lambda}_{\rho\sigma}(A)
M^{\sigma}_{\;\alpha}(A)M^{\rho}_{\;\beta}(A)
(-1)^{\varepsilon_{\alpha}\varepsilon_{\rho}}-\\
\nonumber
&&-\big((M^{-1})^{\gamma}_{\;\mu}(A)\overleftarrow{\pa}_{\!\!A^j}\big)
{\bf R}^j_{\;\alpha}(A)M^{\mu}_{\;\beta}(A)(-1)^{\varepsilon_{\alpha}\varepsilon_{\mu}}+\\
\nonumber
&&+\big((M^{-1})^{\gamma}_{\;\mu}(A)\overleftarrow{\pa}_{\!\!A^j}\big)
{\bf R}^j_{\;\beta}(A)M^{\mu}_{\;\alpha}(A)
(-1)^{\varepsilon_{\alpha}(\varepsilon_{\beta}+\varepsilon_{\mu})}.
\eeq
The relation (\ref{gabold}) is the final form of the gauge
algebra for deformed generators (\ref{generators}). This algebra
is irreducible and closed and includes the deformed structure coefficients
(\ref{newstructure}).
It is interesting to point out that even if the initial gauge theory is Abelian,
the deformed
gauge theory is non-Abelian.

As a result, we have completely built a deformed gauge theory that
is given by the deformed action and the deformed generators. The deformed
generators (\ref{generators}) includes the matrices $M(A)^{i}_{\;j}
= \delta^{i}_{\;j}(A) + h^{i}_{\;j}(A)$ and
$M^{\alpha}_{\;\beta}(A)=
\delta^{\alpha}_{\;\beta}+g^{\alpha}_{\;\beta}(A).$ As we saw, the arbitrary functions
$h^{i}_{\;j}(A)$ define the deformation of the initial action.
Apart from the function $h^{i}_{\;j}(A)$, the
deformed generators are also defined by the arbitrary functions $g^{\alpha}_{\;\beta}(A)$.
Therefore we
can conclude that the arbitrary deformation in the initial gauge theory is defined
by two (in general,
non-local) arbitrary generating functions of the initial fields $A^i.$

In some cases, one can expect that only one of these arbitrary functions will be independent.
As such an example, we consider the form of the deformed generators
in the first order in functions
$h$ and $g$. The general relation (\ref{generators}) leads to
\beq
\label{Rboldap}
{\bf R}^i_{\alpha}(A)=R^i_{\alpha}(A)-
\big(h^i(A)\overleftarrow{\pa}_{\!\!A^j}\big)R^j_{\alpha}(A)+
\big(R^i_{\alpha}(A)\overleftarrow{\pa}_{\!\!A^j}\big)h^j(A)+
R^i_{\beta}(A)g^{\beta}_{\;\alpha}(A)+\cdots ,
\eeq
where $\cdots$ means higher terms in generating functions $h$ and $g$.
Let us assume that the
following equation is fulfilled
\beq
\label{special}
R^i_{\beta}(A)g^{\beta}_{\;\alpha}(A) =
\big(h^i(A)\overleftarrow{\pa}_{\!\!A^j}\big)R^j_{\alpha}(A).
\eeq
Then we have a relation between the functions $g$ and $h$ and only one
of these functions becomes independent.

\section{The Yang-Mills theory as the deformation of the Abelian gauge theory}
\noindent
In this section, we will demonstrate how the general theory
under consideration allows to obtain the non-Abelian Yang-Mills
theory as the deformation of the Abelian gauge theory. To be more
precise, we will show that such a non-local deformed theory possesses
the closed local sector.

We begin with
a free vector field model with the action
\beq
\label{YM0}
S_0[A]=-\frac{1}{4} F^a_{0\mu\nu}(A)F^{a\mu\nu}_0(A),
\eeq
where
\beq
F^a_{0\mu\nu}(A)=\pa_{\mu}A^a_{\nu}-\pa_{\nu}A^a_{\mu},
\eeq
and $a$ is the index of some semi-simple Lie algebra with the structure constants $f^{abc}$.
The action $S_0[A]$ is invariant,
\beq
\delta_{\xi}S_0[A]=0,
\eeq
under the Abelian gauge transformations,
\beq
\delta_{\xi}A^a_{\mu}=\pa_{\mu}\xi^a,\quad R^{ab}_{\mu}=\delta^{ab}\pa_{\mu}.
\eeq
Here $\xi^a$ are arbitrary fields of space-time coordinates $x$ and
$R^{ab}_{\mu}$ are generators of initial gauge symmetry.

According to the general deformation procedure described in Section 3, the deformed action is
given by the relation
\beq
\nonumber
S[A]&=&-\frac{1}{4} F^a_{0\mu\nu}(A+h(A))F^{a\mu\nu}_0(A+h(A))=\\
\nonumber
&=&-\frac{1}{4} F^a_{0\mu\nu}(A))F^{a\mu\nu}_0(A))
-\frac{1}{2} F^a_{0\mu\nu}(A))F^{a\mu\nu}_0(h(A))-\\
&& -\frac{1}{4} F^a_{0\mu\nu}(h(A))F^{a\mu\nu}_0(h(A)),
\label{SAh}
\eeq
where $h(A)=\{h^a_{\mu}(A)\}$ is non-local function responsible
for the deformation of initial action (\ref{YM0}). Using integration
by parts we obtain the following expressions for the two last terms in
the r.h.s of (\ref{SAh}),
\beq
\frac{1}{2}
F^a_{0\mu\nu}(A))F^{a\mu\nu}_0(h(A))=-A^a_{\mu}\Box h^{a\mu}(A)+
A^a_{\mu}\pa_{\nu}\pa^{\mu}h^{a\nu}(A), \eeq \beq \frac{1}{4}
F^a_{0\mu\nu}(h(A))F^{a\mu\nu}_0(h(A))=- \frac{1}{2}h^a_{\mu}(A)\Box
h^{a\mu}(A)  + \frac{1}{2}h^a_{\mu}(A)\pa_{\nu}\pa^{\mu}h^{a\nu}(A)
\eeq

The functions  $h^a_{\mu}(A)$ have the same
indices as the fields $A^a_{\mu}$. Having in mind this fact, one considers
the following function $h^a_{\mu}(A)$
\beq
\label{ha}
h^a_{\mu}(A)=\frac{1}{\Box}\big[c_1\pa^{\nu}(f^{abc}A^b_{\nu}A^c_{\mu})+
c_2f^{abc}f^{cmn}A^{b\nu}A^m_{\nu}A^n_{\mu}\big],
\eeq
where $c_1, c_2$ are some constants. Then, after some transformations, we obtain
\beq
\nonumber
&&\frac{1}{2} F^a_{0\mu\nu}(A))F^{a\mu\nu}_0(h(A))=
\frac{1}{2}c_1F_{0\mu\nu}^{a}(A)f^{abc}A^{b\mu}A^{c\nu}+
c_2f^{abc}A^{b}_{\mu}A^{c}_{\nu}f^{amn}A^{m\mu}A^{n\nu}+\\
&&\qquad\qquad+
\textit{non-local terms containing the $\frac{1}{\Box}$}.
\eeq
Taking the constants in the form $c_1=1, c_2=\frac{1}{4}$ we get
\beq
S[A]=S_{YM}[A]+S_1[A],
\eeq
where
\beq
S_{YM}[A]=-\frac{1}{4}F^a_{\mu\nu}(A)F^{a\mu\nu}(A),\quad
F^a_{\mu\nu}(A)=\pa_{\mu}A^a_{\nu}-\pa_{\nu}A^a_{\mu}+f^{abc}A^b_{\mu}A^c_{\nu}
\eeq
is the standard local Yang-Mills action and $S_1[A]$ is some correction responsible for
non-local deformed action. As a result, we see that the deformation procedure,
described in Section 3,
allows us to derive the action of non-Abelian Yang-Mills theory beginning with
free Abelian gauge theory.

The next step is to get the Yang-Mills gauge transformations.
Taking into account the explicit expression for generating functions
$h^a_{\mu}(A)$ (\ref{ha}) with fixed constants and using the relation
(\ref{special}) one obtains
\beq
{\bf R}^{ab}_{\mu}(A)=D^{ab}_{\mu}(A)+R^{ab}_{1\mu}(A),
\eeq
where $D^{ab}_{\mu}(A)$ is covariant derivative with respect to $A^a_{\mu}$,
\beq
D^{ab}_{\mu}(A)=\delta^{ab}\pa_{\mu}+f^{acb}A^c_{\mu},
\eeq
where $D^{ab}_{\mu}(A)$ are the standard Yang-Mills generators and $R^{ab}_{1\mu}(A)$ corresponds
to the part of  gauge generators responsible for gauge invariance of the complete non-local
deformed action. Thus, in the case under consideration, the non-local deformed
action contains closed local sector which is nothing more than the Yang-Mills theory.
It is interesting to point out that in the given case, the relation (\ref{special})
allows to eliminate all the terms in deformed gauge transformations which contain the
space-time derivatives of the gauge parameters besides $\partial_{\mu}\xi^a$.

\section{Cubic interaction vertex for massless integer higher spin fields as the deformation
of the free massless higher spin theory}
\noindent
In this section, we will show how the approach to the construction of the gauge invariant
deformation allows to derive the cubic interaction vertex in
higher spin field theory.

We begin with free massless integer higher spin $s$ field theory which
is described by the Fronsdal action \cite{Fronsdal-1}
\beq
\nonumber
&&\!\!\!S^{(2)}[\phi]= \!\!\int \!\! d^4x \{-\varphi_{\mu_1...\mu_s}\Box
\varphi^{\mu_1...\mu_s}-
\frac{s}{2}\partial_{\alpha}\varphi^{\alpha\mu_2...\mu_s}
\partial^{\beta}\varphi_{\beta\mu_1...\mu_s}-
\frac{s(s-1)}{2}\varphi^{\rho}{}_{\rho\mu_3...\mu_s}\partial_{\alpha}\partial_{\beta}
\varphi^{\alpha\beta\mu_3...\mu_s}-\\
\label{Fronsdal}
&&\qquad-\frac{s(s-1)}{4}
\partial_{\alpha}\varphi^{\rho}{}_{\rho\mu_3...\mu_s}
\partial^{\alpha}\varphi_{\sigma}{}^{\sigma\mu_3...\mu_s}-
\frac{s(s-1)(s-2)}{8}\partial_{\alpha}\varphi_{\rho}{}^{\rho\alpha\mu_4...\mu_s}
\partial^{\beta}\varphi^{\sigma}{}_{\sigma\beta\mu_4...\mu_s} \}.
\eeq
Here $\varphi_{\mu_1...\mu_s}$ is a totally symmetric double
traceless field with standard bosonic field dimension. The theory is
invariant under the Abelian gauge transformations with the
traceless totally symmetric parameters $\xi_{\mu_1...\mu_{s-1}}.$ We
are going to apply our procedure of deformation to the free theory
with action(\ref{Fronsdal}). However, a point needs to be made here.
The deformation theory under consideration assumes that the fields
and the gauge parameters are completely unconstrained. But the
higher spin fields in action (\ref{Fronsdal}) and the corresponding
gauge parameters obey the traceless constraints and our approach can
not be applied in literal form. To simplify the situation we will
follow the reasoning accepted in the higher spin field theory when
constructing the interaction vertices (see e.g. \cite{metsaev} and
the references therein). Namely, let us assume that there are no
traceless restrictions on the fields and parameters from the very
beginning and then impose needed restrictions on the fields in the
vertices afterwards\footnote{In principle there is a formulation of
free higher spin field theory where the fields in action are
completely unconstrained and the true constraints appear only on
equations of motion \cite{BGK}, \cite{BG}. }.

According to the general procedure, described in Section 3, to obtain
the deformed action we should replace the field
$\varphi_{\mu_1...\mu_s}$ in the action (\ref{Fronsdal}) by the
field $\varphi_{\mu_1...\mu_s}+h_{\mu_1...\mu_s}$ with the arbitrary
totally symmetric non-local function $h_{\mu_1...\mu_s}(\varphi).$
We will show that this function can be taken in such a way that the
deformed theory contains at least cubic interaction local
vertex\footnote{We emphasize that at an arbitrary choice of the
generating function $h$ the deformed action can not include the
closed local sector. This can only be with a very special choice of
the generating function, if even possible at all.}.

Let us consider the function $h_{\mu_1...\mu_s}(\varphi)$ in the
form \beq \label{hs2k}
h^{\mu_1\cdots\mu_{s}}=c_kg^{s+2k}\frac{1}{\Box}
\pa^{\{\mu_1}\cdots\pa^{\mu_k}\varphi^{{\mu_{k+1}}\cdots{\mu_s}\}\nu_1\cdots\nu_k}
\pa_{\nu_1}\cdots \pa_{\nu_k}\pa^{\lambda_1}\cdots\pa^{\lambda_s}
\varphi_{\lambda_1\cdots\lambda_s}. \eeq Here the parameter $k$
takes the values $k=0,1, \ldots s$, $c_k$ are the arbitrary real
constants and $g$ is a coupling constant of the dimension ${\rm
dim}(g)=-1$. The used symbol $\{\mu_1\cdots \mu_k\mu_{k+1}\cdots
\mu_s\}$ means symmetrization with respect to indexes included. The
function (\ref{hs2k}) is the only admissible function generating
cubic vertex with help of transformation $\varphi \rightarrow
\varphi + h$ in the action (\ref{Fronsdal}) and preserving the
symmetry and dimension of $\varphi$. It is clear that in this case,
we need a dimensional coupling constant. Transformation $\varphi
\rightarrow \varphi + h$ in the first term of action
(\ref{Fronsdal}) yields the family of the local cubic vertices \beq
\label{cubic} S^{(3)}_{local}[\varphi] = -2c_kg^{s+2k}\int d^4x
\varphi^{\mu_1\cdots\mu_s}
\pa_{\{\mu_1}\cdots\pa_{\mu_k}\varphi_{{\mu_{k+1}\cdots\mu_s}\}\nu_1\cdots\nu_k}
\pa^{\nu_1}\cdots \pa^{\nu_k}\pa_{\lambda_1}\cdots\pa_{\lambda_s}
\varphi^{\lambda_1\cdots\lambda_s} \eeq with $k=0,1 \ldots s $.
Deformation of the other terms in free action leads to non-local
contributions. The obtained cubic vertices (\ref{cubic}) correspond
to the cubic vertices that were constructed in the papers on higher
spin field theory by different methods (see e.g. the recent papers
\cite{metsaev-1}, \cite{manvelyan}, \cite{metsaev}, \cite{zinoviev}
and the references therein). In our approach, these vertices are a
simple consequence of the general procedure, described in Section 3.

The next question is finding the true local gauge transformations.
Just as in deriving gauge transformations in Yang-Mills theory, we
start with gauge transformations of free theory and apply the
general relations (\ref{Rboldap}). As a result, we obtain the
non-local gauge transformations, corresponding to general non-local
deformed theory. However, it is easy to see that these
transformations contain a local piece, leaving
invariant the action $S^{(2)}+S^{(3)}_{local}$ up to fourth order
terms in $\varphi.$ We see that the theory under consideration
admits a closed local sector. Thus, the general deformation
procedure allows to comparatively simply to describe the cubic
vertices in the massless higher integer spin theory.

\section{Conclusion}
Let us summarize the results. We have described a general procedure
of gauge-invariant deformation of classical gauge theories. The
procedure is based on the use of the BV-formalism where
the central objects are antibracket and the master equation. The
arbitrary gauge deformation is formulated in terms of anticanonical
transformation leaving invariant the antibracket. We have proved
that the arbitrary gauge deformation of a given gauge-invariant
theory is described by two arbitrary generating functions, where one
is responsible for the deformation of the initial action and the
other for the deformation of the gauge generators. The deformations
are realized in the explicit form and given by the relations
(\ref{dAc}) and (\ref{generators}).

The deformation of initial action has an extremely simple form and
means replacing the gauge field $ \phi $ in the initial action
by the field $ \phi + h(\phi) $
with an arbitrary nonlocal generating function $ h(\phi). $
The deformation of gauge generators also has a
simple enough form and is described by the same function
$h(\phi)$ and another arbitrary function $g(\phi).$ We have
calculated the algebra of deformed generators in the form
(\ref{gabold}) with deformed structure coefficients
(\ref{newstructure}). We emphasize that even if the initial theory
is Abelian, the corresponding deformed theory is non-Abelian.

The essential feature of the obtained deformed theory is that it is
non-local in general. However, in special cases, there can be a
situation when such a non-local theory contains a closed local
sector. It means that the deformed action can involve the local
piece which is invariant under the local piece of deformed gauge
transformations. As a result, we can obtain some new local gauge
theory.

The first important test for the theory under consideration is a
possibility to derive Yang-Mills theory. We have shown that if to
start with Abelian gauge theory and apply the transformation $A
\rightarrow A+h(A)$, where the function $h^{a}_{\mu}(A)$ is given by
(\ref{ha}) we obtain some non-local vector field theory with a local
piece in action which just is the Yang-Mills action. The corresponding
piece in the deformed gauge transformations is the Yang-Mills gauge
transformation.

Also, we have considered a derivation of the massless integer higher
spin cubic vertex in the framework of the general deformation
theory. We again started with free theory and constructed in
explicit form the non-local field deformation which generates the
cubic interaction vertex consistent with ones in the literature. The
generating function $h(A)$ should be nonlocal to reproduce
nontrivial deformation of the free action. In the case under
consideration, the non-locality is due to the presence of the
operator $\frac{1}{\Box}$ in the function $h$. To get the
higher orders vertices, we have to consider the non-local
contributions to this function containing e.g. the operators
$\frac{1}{({\Box})^n}, \, n=1,2,3 \ldots .$ Therefore, one can
expect that apparently the interaction vertices of the quartic and
higher orders will obligatorily be non-local\footnote{Non-local contributions
to fourth
order vertex arise already at substituting the
$\varphi \rightarrow \varphi + h$ with
$h$ given by (\ref{hs2k}) to (\ref{Fronsdal}).}. This basically
corresponds to results of the works \cite{Tsulaia-1},
\cite{Taronna}, \cite{Tsulaia}, \cite{Taronna-1}, \cite{Tseytlin},
\cite{Ponomarev}.
We guess that the aspects of the locality of the vertices in the higher
spin field theory deserve a comprehensive study (see e.g.
\cite{Vasiliev} and the references therein).

Let us note some areas of further development and application of our
deformation theory. First, we can generalize the approach for the
theories with dependent generators. Second, it would be useful to
study the relations of the quantum effective actions for the classical
theories obtained one from another by non-local gauge-invariant
deformation. Third, it would also be interesting to explore which
nonlocal theories of gravity can be constructed by the deformation
of the free massless symmetric second rank tensor field theory and
whether there exists in deformed gravity theories a closed local
sector corresponding to Einstein's gravity.

We believe that the most interesting applications of the developed
deformation theory relate to the higher spin field theory. Since the
true cubic interaction vertex for massless integer higher spin
fields was derived on the base of deformation theory, one can expect
that the other vertices for massless and massive higher spin fields
can also be derived in the framework of such an approach. We plan to
study all these issues in the forthcoming works.

\section*{Acknowledgments}
\noindent
The authors are grateful to M. Tsulaia for the  discussion of
the higher spin vertices. The work is supported
by the Ministry of Education of the Russian Federation, project
FEWF-2020-0003.

\end{document}